\documentclass{aastex}
\usepackage{emulateapj5}

\slugcomment{accepted by AJ}

\shortauthors{A. K. Inoue}
\shorttitle{Lyman Continuum Extinction}

\begin{document}

\title{Lyman Continuum Extinction by Dust in H {\sc ii} Regions of Galaxies}

\author{AKIO K. INOUE} 

\affil{Department of Astronomy, Faculty of Science, Kyoto University,
Sakyo-ku, Kyoto 606-8502, JAPAN}
\email{AKI: inoue@kusastro.kyoto-u.ac.jp}

\begin{abstract}

We examine Lyman continuum extinction (LCE) in H II regions 
by comparing infrared fluxes of 49 H II regions in the Galaxy,
M31, M33, and the LMC with estimated production rates of Lyman
continuum photons.  A typical fraction of Lyman continuum photons
that contribute to hydrogen ionization in the H II regions of three
spiral galaxies is $\la$ 50 \%.  The fraction may become smaller as the
metallicity (or dust-to-gas ratio) increases.  
We examine the LCE effect on estimated star formation rates (SFR) 
of galaxies.  The correction factor for the Galactic dust-to-gas 
ratio is 2-5.

\end{abstract}


\section{Introduction}

Dust grains exist everywhere ---
from circumstellar environment to interstellar space.
Hence, radiation from celestial objects is always absorbed and
scattered by dust grains.
Without correction for the dust extinction, we inevitably underestimate
the intrinsic intensity of the radiation, and our understanding of
physics of these objects may be misled.
Thus, dust extinction for nonionizing photons ($\lambda > 912$ \AA)
in the interstellar medium (ISM) has been well studied to date (e.g.,
interstellar extinction curve: \citealt{sav79, sea79, cal94, gor00}).

In H {\sc ii} regions, Lyman continuum (LC) photons also 
suffer extinction by dust grains (e.g., \citealt{ish68, har71}).
We refer to this another type of extinction as the Lyman continuum
extinction (LCE).
Indeed, the fraction of LC photons absorbed by dust in
H {\sc ii} regions can be large (e.g., Petrosian, Silk, \& Field 1972;
\citealt{pan74, mez74, nat76, sar77, smi78, aan89, shi95, bot98}).
\cite{pet72}, for example, have estimated the fraction of LC
photons contributing to hydrogen ionization to be 0.26 in the Orion
nebula.

If a large amount of LC photons is absorbed by dust in H
{\sc ii} regions, the effect of the LCE on estimating the star formation 
rate (SFR) will be very large.
This is because we can only estimate the number of ionizing photons, LC
photons used for hydrogen ionization, from the observation of hydrogen
recombination lines or thermal radio continuum.
When we define the fraction of ionizing photons estimated from
observations as $f$, we obtain 
\begin{equation}
 N'_{\rm LC} = f N_{\rm LC}\,,
 \label{eq1}
\end{equation}
where $N_{\rm LC}$ and $N'_{\rm LC}$ are the intrinsic and apparent
production rates of LC photons, respectively.
Unless we correct the observed data for the LCE, we cannot obtain the
actual LC photon production rate, and then, the true SFR.
Therefore, we should estimate the fraction, $f$, in H {\sc ii} regions.
However, the LCE has not been discussed in the context of
estimating the SFR of galaxies before Inoue, Hirashita, \& Kamaya
(2001a, hereafter Paper I).

Paper I has proposed two independent methods for estimating $f$, and
applied the methods to the Galactic H {\sc ii} regions.
Then, they have shown that there are a number of H {\sc ii} regions with
$f \la 0.5$ in the Galaxy.
Moreover, they have found that $f$ can be approximated to be only a
function of dust-to-gas ratio.
If the approximation can be applied to the nearby spiral galaxies and
their observed dust-to-gas ratios are adopted, $f$
averaged over galactic scale is about 0.3. 
Therefore, the SFR corrected by the LCE effect is
3 times larger than the uncorrected SFR.

The aim of this paper is that we confirm these results of Paper I not
only for H {\sc ii} regions in our Galaxy but for extragalactic H {\sc
ii} regions.
We also discuss if the effect of the LCE on estimating the SFR of
galaxies is really important.
We describe the method for estimating $f$ in section 2.
Then, $f$ is determined for the H {\sc ii} regions in the local group
galaxies in section 3.
We discuss the dependence of $f$ on the dust-to-gas ratio in
section 4, and our conclusions are summarized in the final section.

\section{Method for Estimating the Effect of LCE}

To estimate $f$ in equation (\ref{eq1}) for each H {\sc ii} region, the
apparent production rate of LC photons in each region, $N'_{\rm LC}$, is
compared to its IR luminosity, $L_{\rm IR}$.
In this paper, we use a method for estimating $f$ presented in section 3 
of Paper I.

According to equation (12) in Paper I, we obtain
\begin{equation}
 f = \frac{0.44 + 0.56\epsilon}
          {0.28 + 5.6(L_{\rm IR,7}/N'_{\rm LC,50})}\,,
 \label{eq2-1}
\end{equation}
where $\epsilon$ denotes the average efficiency of dust absorption for
nonionizing ($\lambda > 912$ \AA) photons, and $L_{\rm IR,7}$ and $N'_{\rm
LC,50}$ are the IR luminosity and LC photon production rate normalized
by $10^7$ $L_\sun$ and $10^{50}$ s$^{-1}$, respectively.
The equation is based on the theory of the IR emission from H {\sc
ii} regions by \cite{pet72} and the following discussion by
\cite{ino00}\footnote{See also \cite{ino1b}}.
It is also worthwhile to note that $L_{\rm IR,7}/N'_{\rm LC,50}$ is
proportional to the flux ratio of the IR emission to the thermal radio
emission, H$\alpha$ line, etc.
If we express in the form of the flux ratio, the determination of $f$ is
free from the uncertainty of the distance to the object.

Here, we describe some assumptions adopted in the derivation of equation 
(\ref{eq2-1}).
Since $L_{\rm IR}$ is the total energy emitted by dust, the wavelength
range is set to 8--1000 \micron, which covers almost all wavelength
range of dust emission.
Also, it is supposed that the IR radiation from H {\sc ii}
regions is dominated by large ($\sim 0.1 \micron$) grains in thermal
equilibrium with ambient radiation field.
In other words, we neglect the contribution to the total IR luminosity
of very small grains ($\lesssim 0.01 \micron$) and polycyclic
aromatic hydrocarbons (PAHs), which emit mainly in the mid-infrared
(MIR, 8--40 \micron) (e.g., \citealt{dwe97}).
In principle, the assumption would make us underestimate the total IR
luminosity of dust in H {\sc ii} regions because such MIR emission is
observed from these regions (e.g., \citealt{pis91, deg92, lei98}).
Fortunately, it seems that the contribution of the MIR luminosity to
that of the total IR is enough small to be neglected (Figure 1 in
\citealt{lei98}).
Thus, the IR spectral energy distribution (SED) of H {\sc ii} regions is 
assumed to be a modified black-body function with one temperature
through the paper.

Moreover, we do not take account of the existence of helium.
This is because almost all (96\% according to \citealt{mat71})
photons produced by helium recombination can ionize neutral hydrogen.
We do not consider the escape of LC photons from H {\sc ii} regions,
either.
Indeed, less than 3 \% of LC photons can escape from the star-forming
regions in nearby galaxies \citep{lei95}.
In short, we deal with an ideal H {\sc ii} region that consists of only
hydrogen and dust, and where no LC photons leak.

Let us determine $\epsilon$, which denotes the average efficiency of
dust absorption for nonionizing photons.
The total (absorption and scattering) optical depth of dust,
$\tau_\lambda$, is expressed by 
\begin{equation}
 \tau_\lambda = \frac{\tau^{\rm abs}_\lambda}{1-\omega_\lambda}
              = \frac{A_\lambda}{2.5\log e}\,,
 \label{eq2-2}
\end{equation}
where $\tau^{\rm abs}_\lambda$, $\omega_\lambda$, and $A_\lambda$ are the
absorption optical depth of dust, the dust albedo, and the amount of
extinction at wavelength, $\lambda$, respectively.
We can determine $A_\lambda$ from an extinction curve, $X_\lambda$, and
its normalization, $E_{B-V}$, i.e. $X_\lambda = A_\lambda / E_{B-V}$.
The efficiency of dust absorption at wavelength, $\lambda$,
$\epsilon_\lambda$ is defined as 
\begin{equation}
 \epsilon_\lambda \equiv 1 - e^{-\tau^{\rm abs}_\lambda}
  = 1 - 10^{-0.4 X_\lambda E_{B-V} (1-\omega_\lambda)}\,.
 \label{eq2-3}
\end{equation}
Then, averaging $\epsilon_\lambda$ over $\lambda$ 
with weight of the intrinsic stellar flux density yields the average
efficiency, $\epsilon$.

\begin{figure}
\plotone{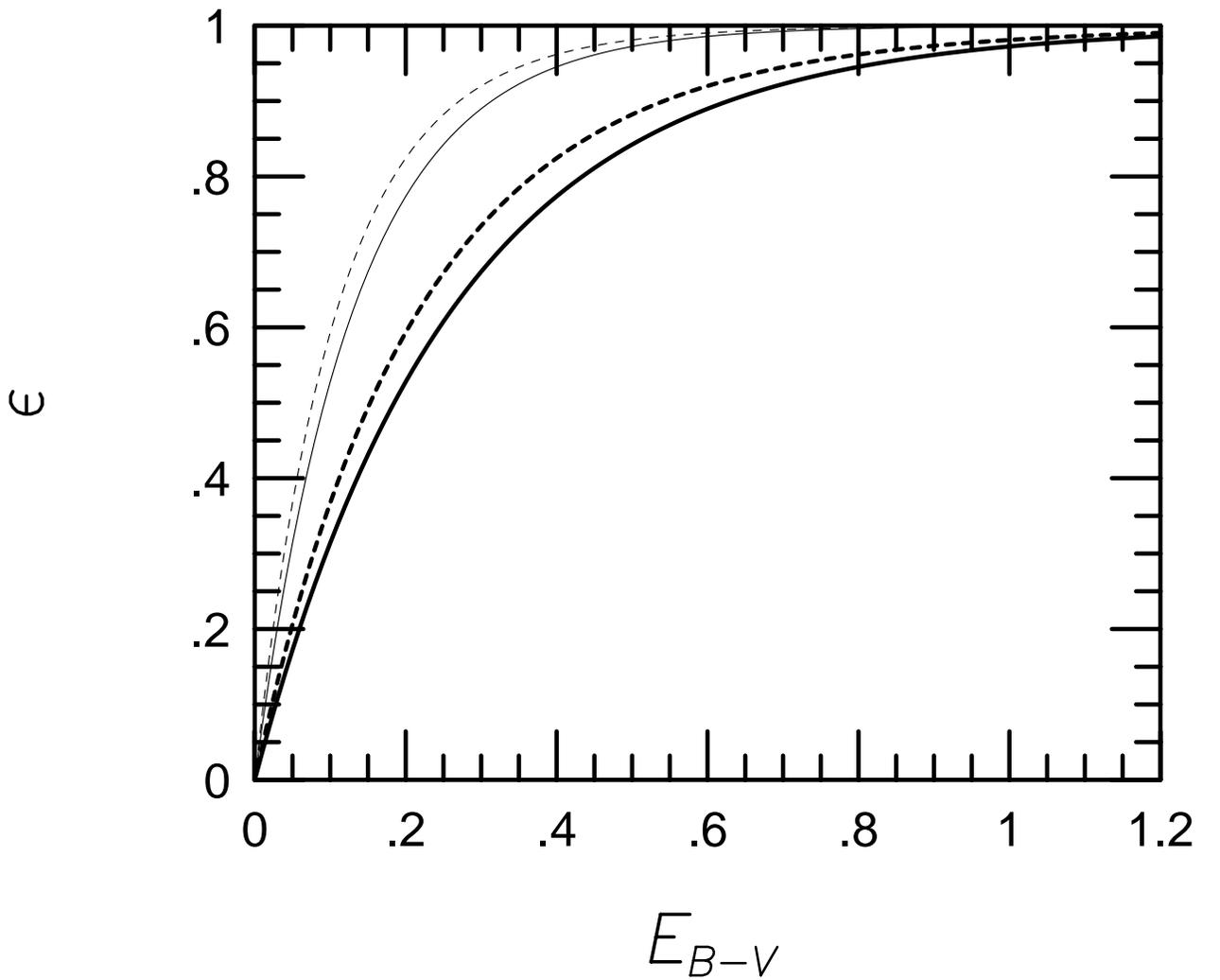}
\figcaption{Average extinction efficiency for nonionizing
 ultraviolet photons, $\epsilon$, as a function of color excess,
 $E_{B-V}$. The solid lines are made by using the Galactic extinction
 curve \citep{sea79}. The LMC extinction curve \citep{how83} results in the
 dashed lines. The thick lines denote the results of cases that dust
 albedo is set to 0.5. The thin lines are the results of no
 scattering. \label{f1}}
\end{figure}

We show $\epsilon$ as a function of $E_{B-V}$ in Figure \ref{f1}.
In this calculation, we set the stellar spectrum to be simply the
black-body of 30000 K.
The change of the temperature dose not alter the result significantly.
For example, if we choose 50000 K as the effective temperature of stars, 
the value of $\epsilon$ increases with a factor of about 1 \%.
The calculation of the average is performed from 1100 to 3500 \AA.
We adopt two types of the interstellar extinction curve, $X_\lambda$:
one is that in our Galaxy \citep{sea79}, the other is that in the Large
Magellanic Cloud (LMC) \citep{how83}.
Each case is shown in Figure \ref{f1} by the solid and dashed lines,
respectively.
Moreover, we set the dust albedo to be a constant of 0.5 for
simplicity, although it varies with wavelength \citep{wit73, lil76}.
The results in this case are shown by the thick lines.
Also, the results without scattering (i.e. $\omega_\lambda = 0$) are
shown by the thin lines for comparison.

Once we determine $\epsilon$ for H {\sc ii} regions from Figure
\ref{f1}, we can estimate $f$ from the ratio of the IR luminosity to the
observed LC photon production rate via equation (\ref{eq2-1}).
In the next section, we will examine $f$ for H {\sc ii} regions in some
Local Group galaxies.

\section{Fraction $f$ of H {\sc ii} Regions in the Local Group Galaxies}

In order to determine $f$, the data of the IR luminosity and the
production rate of LC photons for the individual H {\sc ii} region are
required.
But, when we compare two photometric data, we must match their apertures
and resolutions with each other, because, in general, different
observations have different aperture and resolution.
It prevents us from constructing uniform data of a large number of H
{\sc ii} regions.
These effects produce one of the largest uncertainties in
the current analysis.

We have gathered the data that allow us to estimate $f$ with proper
accuracy, though the sample size is rather small.
In this section, we estimate $f$ of H {\sc ii} regions in our Galaxy,
M31, M33, and LMC.

\subsection{Our Galaxy}

Suitable data for our analysis have been compiled by \cite{wyn74}.
The data are read out from their figure 9 directly, and are tabulated in
Table \ref{t1}.
The column 1 is the name of H {\sc ii} regions.
Since two separate components of W3, W51, and NGC 6537 are plotted in
figure 9 of \cite{wyn74}, we list both components of these regions in
Table \ref{t1}.
The columns 2 and 3 are the logarithmic production rate of LC photons
estimated from radio observations\footnote{The estimated production rate
of LC photons of M8, Orion, and M17 in this paper differ from those in
Table 1 of Paper I because their references are different from those
used in this study.} and FIR (40--350 \micron) luminosity, respectively.
The FIR (40--350 \micron) luminosity is almost equal to the total IR
(8--1000 \micron) luminosity if the dust SED is assumed to be the
modified black-body function with the temperature of 30 K and the
emissivity index of 1.

Since $E_{B-V} \sim 1$ for the Galactic H {\sc ii} regions (e.g., 
\citealt{cap00}), Figure \ref{f1} tells us that $\epsilon \simeq 1$,
even in the case with the scattering for nonionizing photons.
Thus, we can safely assume $\epsilon = 1$ for the Galactic H {\sc ii}
regions, so that we can determine $f$ of these regions via equation
(\ref{eq2-1}) (column 5).
The mean value of $f$ is 0.45.
Thus, we find that about half of LC photons is absorbed by
dust in H {\sc ii} regions.
This is one of the results in Paper I.

\subsection{M31}

\cite{xu96} have studied the properties of discrete FIR sources 
in M31 by using the {\it IRAS} HiRes image (\citealt{aum90, ric93}).
Thirty nine sources are extracted from the 60 \micron\ map.
They have also examined the correlation between these FIR sources and
H$\alpha$ sources presented by \cite{wal92}.
Since \cite{wal92} have surveyed only the northeast half of M31, 12 out
of 39 FIR sources have H$\alpha$ counterparts.
These twelve FIR sources make the sample for the current analysis.
In Table \ref{t2}, some properties of the sample H {\sc ii} regions in
M31 are listed.

The column 2 of Table \ref{t2} is the measured flux density at 60 \micron.
Unfortunately, only the 60 \micron\ flux densities of almost all sample
regions are available.
Hence, we estimate total IR luminosity of dust emission in each region
from the 60 \micron\ flux density by the following way: 
Assuming the IR SED of these FIR sources to be the modified black-body
function with temperature of 30 K and emissivity index of 1, we obtain 
$L_{\rm IR}/L_\sun = 3.50 \times (D/{\rm kpc})^2 (F_{60\micron}/{\rm
Jy})$, where the wavelength range of the IR luminosity, $L_{\rm IR}$, is
8--1000 \micron, and $D = 760$ kpc is the adopted distance to M31.
The derived IR luminosities are shown in the column 3 of Table \ref{t2}.
The adopted dust temperature of 30 K is reasonable because dust
temperature of 10 isolated H {\sc ii} regions whose flux densities at
both 60 and 100 \micron\ are listed in Table 3 of \cite{xu96} is
estimated to be 29 K on average.

Since the angular resolution of H$\alpha$ image is much higher than that
of 60 \micron\ image, numerous H$\alpha$ sources are included in the
area over which the flux density of one FIR source at 60 \micron\ is
integrated.
The sum of the H$\alpha$ fluxes in the area of each FIR source is
shown in the column 4 of Table \ref{t2}. 
These values are corrected for the [N{\sc ii}] contamination by assuming
a constant ratio of [N{\sc ii}]/H$\alpha$ = 0.3 \citep{wal92}.

When we estimate the production rate of LC photons from H$\alpha$ flux,
we must correct it for the interstellar extinction.
\cite{wal92} have commented that the mode of the interstellar extinction
at H$\alpha$ is 0.8 mag.
Since this value of extinction is estimated from the column density of
H{\sc i} gas, it is a typical extinction averaged over the ISM of M31.
If we assume the Case B and the gas temperature of 10000 K
\citep{ost89}, we obtain $N'_{\rm LC}/{\rm s^{-1}} = 8.77\times 10^{55}
(D/{\rm kpc})^2 (F_{\rm H\alpha}/{\rm ergs\, s^{-1}\, cm^{-2}}) \times
10^{0.4 A_{\rm H\alpha}}$, where $F_{\rm H\alpha}$ is H$\alpha$ flux
(column 4) and $A_{\rm H\alpha}=0.8$ mag is the assumed amount of
extinction at H$\alpha$ line.
The derived $N'_{\rm LC}$ is in the column 5.
The ratio of the IR luminosity to the LC photon production rate is 
shown in the column 6.

Now, we need to estimate $\epsilon$.
Adopting $A_V/E_{B-V}=3.1$ and the Galactic extinction curve, $E_{B-V}$
is $\sim 0.3$ mag when $A_{\rm H\alpha} \sim 0.8$ mag.
Then, we can assume $\epsilon = 0.7$ for the H {\sc ii} regions in M31
from the line for the dust albedo of 0.5 in Figure \ref{f1}.
Finally, we determine $f$ via equation (\ref{eq2-1}) (column 7).
The mean $f$ is 0.38, which is almost equal to that of the Galactic H
{\sc ii} regions.

Also, \cite{dev94} have compared the H$\alpha$ emission map of M31 with
the IR emission map produced by the maximum correlation method
(IRAS HiRes image).
They divided the H$\alpha$ emission into three components; the
star-forming ring, the bright nucleus, and diffuse (unidentified)
component.
The star-forming ring, whose diameter is 1\fdg65, has the observed
H$\alpha$+[N{\sc ii}] luminosity of $4.77\times10^6 L_\sun$.
They estimated its IR (8--1000 \micron) luminosity to be $1.08\times 10^9
L_\sun$ by assuming the modified black-body function with 26 K and the
index of 1.
If we regard the star-forming ring as an aggregation of a lot of
H {\sc ii} regions, and assume $A_{\rm H\alpha}=0.8$ mag and [N{\sc
ii}]/H$\alpha$ = 0.3, we obtain $L_{\rm IR,7}/N'_{\rm
LC,50}\sim 0.5$ for the ring.
This corresponds to $f \sim 0.3$ when $\epsilon=0.7$.
The value of $f$ shows a good agreement with the mean $f$ of the sample
in Table \ref{t2}.

\subsection{M33}

We find suitable data of H {\sc ii} regions in M33 in Devereux, Duric,
\& Scowen (1997), who examine correlation between H$\alpha$, FIR, and
thermal radio emission of M33.
They have reported some properties of 8 isolated H {\sc ii} regions.
We determine $f$ of these regions.
The sample properties are tabulated in Table \ref{t3}.

In the column 2, FIR (40--1000 \micron) luminosities of the H {\sc
ii} regions, which are determined from the flux densities at 60 and 100
\micron\ measured from {\it IRAS} HiRes images, are presented.
In the column 3, the 5 GHz flux densities (see also \citealt{dur93}) are
shown, and the estimated LC photon production rates are in the column 4
via $N'_{\rm LC}/{\rm s}^{-1}=8.88 \times 10^{46} (D/{\rm kpc})^2
(F_{\rm 5GHz}/{\rm Jy})$ \citep{con92}.
For the distance to M33, 840 kpc is adopted.
If dust SED can be fitted by modified black-body function (emissivity
index of 1, 30K), we have $L_{\rm FIR}$ (40--1000 \micron) $\sim L_{\rm
IR}$ (8--1000 \micron).
We determine the ratio of the IR luminosity to the photon
production rate under this assumption (column 5).

\cite{dev97} have also commented on the properties of the interstellar
extinction for the H {\sc ii} regions in M33.
They have compared H$\alpha$ emission with thermal radio emission, and
argued in their figure 7 that the average extinction is $A_V \sim 1$ mag
(i.e. $A_{\rm H\alpha} \sim 0.8$ mag), although there is a large
dispersion corresponding to $\pm 1$ mag.
Thus, we can assume $\epsilon = 0.7$ for all the sample regions in M33
as well as those in M31.
The determined values of $f$ are listed in the column 6 of Table
\ref{t3}.
The mean is 0.41, although the large dispersion of the interstellar
extinction may cause a large dispersion of $f$.

Thus, we saw that $f \sim 0.4$ or less for the H {\sc ii} regions
in M31 and M33.
Hence, we conclude that the typical values of $f$ in these
galaxies are nearly equal to that in our Galaxy.
Therefore, we suggest that the fraction of LC photons consumed by dust
in the H {\sc ii} regions of other spiral galaxies may be a $\ga 50 \%$
as well as that in the Galaxy, M31, and M33.

\subsection{LMC}

\cite{deg92} have provided the data of 6 H {\sc ii} regions in the LMC.
The IR flux densities of the sample regions have been measured from the
co-added survey data by {\it IRAS}.
Relatively isolated regions have been chosen as the sample regions in
order to perform a reasonable subtraction of the background flux.
\cite{deg92} have reported the IR, H$\alpha$, and radio fluxes of the
sample regions whose apertures have been matched carefully among images
of different wavelengths.
The properties of the data are shown in Table \ref{t4}.

The columns 2 and 3 are {\it IRAS} 60 and 100 \micron\ flux densities,
respectively.
The IR (8--1000 \micron) luminosity is calculated from these flux
densities via $L_{\rm IR}/L_\sun = 0.61 \times (D/{\rm kpc})^2
[2.58(F_{60}/{\rm Jy})+(F_{100}/{\rm Jy})]$.
Here, we assume the dust SED to be modified black-body function
with 30 K and index of 1, that is, $L_{\rm IR}$ (8--1000 \micron) $\simeq$
1.66 $L_{\rm FIR}$ (40--120 \micron).
Then, the estimated total IR luminosities are shown in the column 4.
The adopted distance to the LMC is 51.8 kpc.
{}From the flux densities at 5 GHz in the column 5, the production rates
of LC photons are calculated via Condon's formula in section 3.3 
(column 6).

The color excess of each region, $E_{B-V}$, estimated from the observed
Balmer decrement and corrected for a foreground reddening is listed in
the column 8.
Using the line of the LMC extinction law and the dust albedo of 0.5 in
Figure \ref{f1}, we obtained the value of $\epsilon$ from each $E_{B-V}$
(column 9).
Finally, $f$ is determined for each sample region (column 10).
The mean $f$ is 0.74.
This is obviously larger than those of the Galaxy, M31, and M33.

\cite{deg92} have also determined $f$ for these regions.
Her mean $f$ is $0.65\pm0.13$, which is consistent with that in the
current paper.
Here, we should note that $f$ of \cite{deg92} is determined from the dust
optical depth for UV (1000--3000 \AA) photons.
By the definition of $f$, however, it should be determined from the dust
optical depth for LC photons ($\lambda < 912$ \AA) such as we stated in
Paper I or this paper.

On the other hand, \cite{xu92} have compared the IR, H$\alpha$, and
radio emission from the LMC, and have examined more global correlations
among these emissions.
They have displayed appealing contour maps of these wavelengths,
which strongly make us believe that these three kinds of emission have
the same origin, i.e. the star-forming regions.
Indeed, there are excellent coincidences among IR, thermal radio, and
H$\alpha$ sources (see also \citealt{mai79,fur87} for the Galactic
sources).
According to \cite{xu92}, the FIR (40--120 \micron) flux associating
with the star formation in the LMC is $28.9\times10^{-10}$ W m$^{-2}$,
and the thermal radio flux density at 6.3 cm in the same area is 131
Jy.\footnote{Their integrated area is not whole of the LMC. See their
description for detail.}
Since the corresponding $L_{\rm IR,7}/N'_{\rm LC}$ is 0.13, $f \sim
0.7$ when we assume $\epsilon = 0.4$ (the average value for the sample in
Table \ref{t4}).
This is in good agreement with the mean $f$ of the sample in Table
\ref{t4}.

Therefore, we conclude that $f$ of the H {\sc ii} regions in the LMC is
about 0.7 and larger than those in the spiral galaxies like the Galaxy,
M31, and M33.
The larger $f$ may be caused by the lower metallicity of the LMC.
A lower metallicity is likely to lead to a smaller dust-to-gas ratio, so 
that the fraction of ionizing photons, $f$, increases.
The issue will be discussed more closely in the next section.

\section{Discussions}

We have estimated $f$ for each H {\sc ii} region in some Local 
Group galaxies by comparing its observed IR luminosity with production
rate of LC photons.
Here, we discuss what the determined $f$ values suggest.

\subsection{Value of $f$ as a function of metallicity/dust-to-gas ratio}

First, we discuss the relationship between the mean $f$ of each galaxy
and its metallicity (or dust-to-gas ratio).
We naturally expect that the effect of the LCE becomes
larger as the dust content increases.
Then, we also expect the mean $f$ of sample H {\sc ii} regions to be a
function of dust-to-gas ratio or metallicity of their host galaxy.
In order to examine the issue, we summarize some parameters of
each galaxy in Table \ref{t5}.

The mean $f$ with its sample standard deviation is shown in the column 2
of Table \ref{t5}.
Also, we find metallicities and dust-to-gas mass ratios of sample
galaxies in the columns 3 and 4, respectively.
Here, the metallicities are taken from \cite{van00}.
The dust-to-gas mass ratios except for that of M31 are taken from
\cite{iss90}, who have estimated the dust-to-gas ratio from the ratio
of the amount of visual extinction to the surface density of H
{\sc i} gas.
For M31, we determined the dust-to-gas ratio from the observed dust mass
and gas mass directly.
According to the observation of \cite{haa98} by {\it ISO}, M31 have $3
\pm 1 \times 10^7 M_\sun$ as the dust mass.
Also, M31 contains H {\sc i} gas of $3.8 \times 10^9 M_\sun$ \citep{cra80}
and H$_2$ gas of $2.7 \times 10^8 M_\sun$ \citep{dam93}.
Thus, the dust-to-gas mass ratio of M31 is $7.4 \times 10^{-3}$.
Moreover, the dust-to-gas ratios of sample galaxies are normalized by a
typical one of the ISM in our Galaxy, $6 \times 10^{-3}$ \citep{spi78}.

\begin{figure}
\plotone{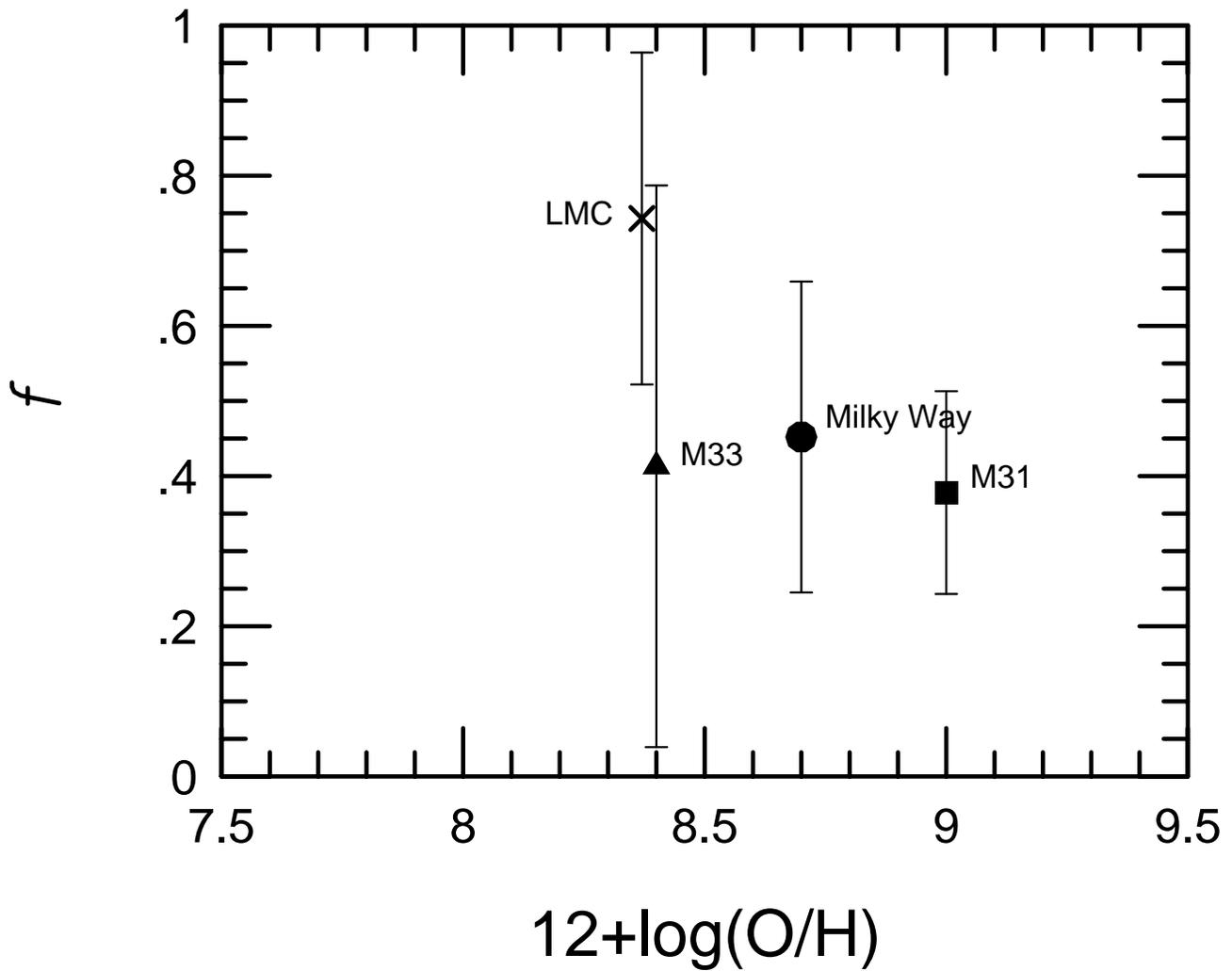}
\figcaption{Fraction of Lyman continuum photons contributing to
hydrogen ionization, $f$, as a function of metallicity. The data of
12+log(O/H) are taken from \cite{van00}. The error bars denote the sample
standard deviation of the mean presented in Table \ref{t5}.\label{f2}}
\end{figure}

\begin{figure}
\plotone{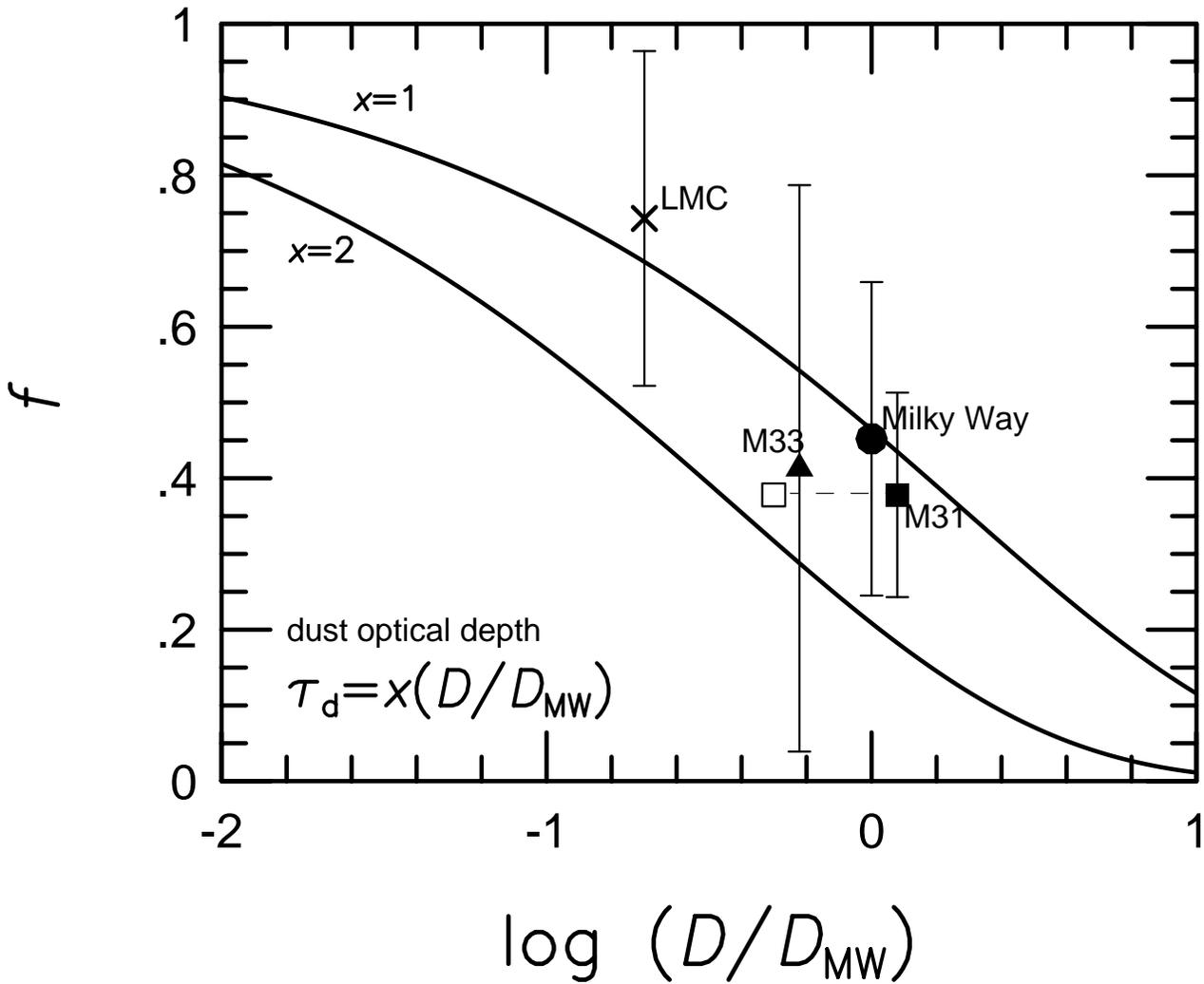}
\figcaption{Fraction of Lyman continuum photons contributing to
 hydrogen ionization, $f$, as a function of dust-to-gas mass ratio,
 $\cal D$. The dust-to-gas ratio is normalized by a typical Galactic
 value, ${\cal D}_{\rm MW}=6\times10^{-3}$ \citep{spi78}. The open
 square represents the case that $\cal D$ of M31 is taken from Issa et
 al.\ (1990). The error bars denotes the sample standard deviation of
 the mean. Two solid lines represent predictions by the model that the
 dust optical depth over the ionized region is proportional to $\cal D$
 like equation (\ref{eq4-1}). \label{f3}}
\end{figure}

In Figure \ref{f2}, mean $f$ values of H {\sc ii} regions in each
galaxy are plotted as a function of metallicity.
We may find a trend that the mean $f$ decreases as the metallicity
increases.
We also show that mean $f$ as a function of dust-to-gas mass ratio in
Figure \ref{f3}.
In this figure, the mean $f$ seems to become small as the
dust-to-gas ratio becomes large.
However, since the number of sample galaxies is still small, more
investigation is needed.
For example, the dependence of $f$ on metallicity will become clearer
if we examine H {\sc ii} regions in the Small Magellanic Cloud or other
metal poor galaxies, or examine the relation of this issue to the
abundance gradient of galaxies.

Let us see how to produce two solid lines in Figure \ref{f3}.
In Parer I, we have formulated the dust optical depth for LC photons,
$\tau_{\rm d}$, as a function of its dust-to-gas ratio (see also
\citealt{hir01});
\begin{equation}
 \tau_{\rm d} = x \left(\frac{\cal D}{\cal D_{\rm MW}}\right)\,,
 \label{eq4-1}
\end{equation}
where $\tau_{\rm d}$ is evaluated over the actual ionized radius of a
spherical H {\sc ii} region containing dust grains uniformly.
Here, we estimate $\tau_{\rm d}$ at the Lyman limit approximately.
In the right hand side, $\cal D$ is the dust-to-gas mass ratio which is
normalized by a typical Galactic ISM value, ${\cal D}_{\rm
MW}=6\times10^{-3}$ \citep{spi78}. 
The coefficient, $x$, is a factor depending on the production rate of LC 
photons and the gas density of the H {\sc ii} region.
For a galaxy having the Galactic dust-to-gas ratio, the
coefficient, $x$, means the dust optical depth for LC photons itself.
We find, in Paper I, $x \sim 2$ for the observed LC photon production
rates and electron number densities in the sample of Galactic H {\sc ii}
regions.\footnote{This sample of Galactic H {\sc ii} regions differs
from the sample in Table \ref{t1} of this paper, and the method
estimating $f$ is also different from that in the paper. See Paper I for
detail.}
\cite{pet72} have derived the relation between the monochromatic dust
optical depth, $\tau_{\rm d}$, and $f$ as the following;
\begin{equation}
  f = \frac{\tau_{\rm d}^3}
     {3\{e^{\tau_{\rm d}}(\tau_{\rm d}^2 - 2\tau_{\rm d} + 2) - 2\}}\,.
 \label{eq4-2}
\end{equation}
The relation is shown in figure 1 of Paper I.
When $\cal D$ is given, therefore, we determine $f$ via equations
(\ref{eq4-1}) and (\ref{eq4-2}).
The data points in Figure \ref{f3} are well reproduced by setting
$x\simeq 1$--2 in equation (\ref{eq4-1}).

Here, we assume that the dust-to-gas ratio in H {\sc ii} regions to be
almost equal to a typical value of that in the ISM of the host
galaxy, although the ratio in local may vary significantly across the
galaxy \citep{sta00}.
On the other hand, the mean value of the dust-to-gas ratio of the H
{\sc ii} gas in our Galaxy is almost equal to those of the H {\sc i} and
H$_2$ gas \citep{sod97}.
Thus, the assumption may be valid globally.

In this paper, the determined $f$ values are somewhat larger than
those in Paper I systematically.
That is, the suitable $x$ parameter in equation (\ref{eq4-1}) for the
data points is 1 rather than 2 which is the best fit value in Paper I.
This may be because we neglect the MIR component emitted by very small
grains and PAHs in the dust IR luminosity (see section 2).
Thus, $f$ determined here may be an upper limit.
Also, $f$ in Paper I is considered to be a lower limit because the
scattering of LC photons by dust is not taken into account in Paper I.

Anyway, the model presented by equations (\ref{eq4-1}) and (\ref{eq4-2}) 
can reproduce the trend of the observational data, when the coefficient, 
$x$, is set to between 1 and 2.
In short, the $\cal D$--$f$ relation falls in the area between two
curves in Figure \ref{f3}.
Therefore, we conclude that $f$ is expected a function of 
dust-to-gas ratio (or metallicity): as the dust content increases, $f$
becomes smaller.

\subsection{Effect of LCE on estimating SFR}

\begin{figure}
\plotone{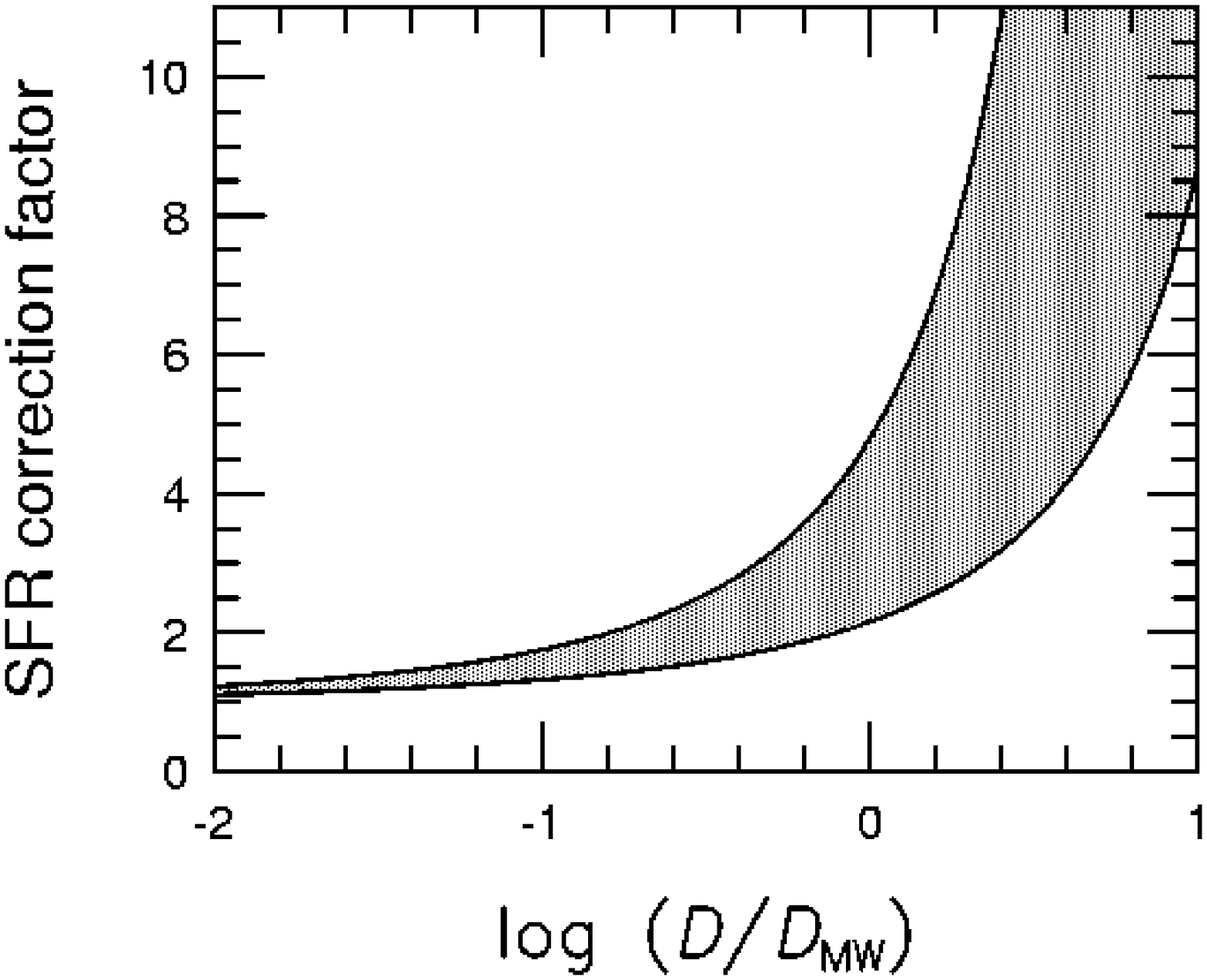}
\figcaption{Correction factor for star formation rate by the LCE
 effect as a function of dust-to-gas ratio. The correction factor is
 $1/f$. The below and above lines are the case of $x=1$ and 2 in
 equation (\ref{eq4-1}), respectively. The correction factor is expected
 in the shaded area between two curves.\label{f4}}
\end{figure}

Finally, we discuss the effect of the LCE on estimating the SFR.
By the definition of $f$ in equation (\ref{eq1}), we obtain the
correction factor for the SFR by the LCE as $1/f$.
If $f$ is a function of dust-to-gas ratio, the correction factor is
also a function of dust-to-gas ratio.
This is shown in Figure \ref{f4}.

Clearly, we find that the correction factor increases as dust-to-gas 
ratio increases.
For nearby spiral galaxies, their dust-to-gas ratios distribute around
the Galactic value (\citealt{alt98, sti00}, see also Paper I).
Indeed, even the starburst galaxies (e.g., M82) and the ultra-luminous
IR (ULIR) galaxies (e.g., Arp 220) are likely to have about the Galactic 
dust-to-gas ratio \citep{kru90, lis00}.
For such dust-to-gas ratio, we expect that the correction factor is 
2--5.
Thus, the effect of the LCE is more important than the uncertainty of
the IMF, which is about a factor of 2 (e.g., Inoue et al.\ 2000).
Therefore, we should take account of the LCE effect when we estimate the
SFR of nearby spiral galaxies.

\section{Conclusions}

To examine the Lyman continuum extinction (LCE) in H {\sc
ii} regions, we compared the observed infrared emission with production
rate of Lyman continuum photons for 49 H {\sc ii} regions in our Galaxy,
M31, M33, and LMC.
Then, we estimate the fraction of Lyman continuum photons contributing
to hydrogen ionization, $f$, in these regions.
We reached the following conclusions:

[1] In many H {\sc ii} regions, $f$ is smaller than 0.5.
The mean $f$ of the sample regions in spiral galaxies (our Galaxy, M31,
and M33) is about 0.4.
On the other hand, the mean $f$ in the H {\sc ii} regions in the LMC is
about 0.7.

[2] The mean $f$ of sample H {\sc ii} regions in each galaxy may be a
function of metallicity or dust-to-gas ratio of their host galaxy.
Namely, $f$ decreases as the metallicity or dust-to-gas ratio increases.

[3] The observational trend is reproduced very well by the model
that the dust optical depth for Lyman continuum photons over the ionized 
region is proportional to the dust-to-gas ratio.
The dust optical depth is in between 1 and 2 for the Galactic
dust-to-gas ratio.

[4] It is expected that the correction factor for the star formation
rate by the LCE increases as dust-to-gas ratio (metallicity)
increases.
The expected correction factor for the galaxies with the Galactic
dust-to-gas ratio is 2--5.
Therefore, the LCE effect should be taken into account when we
estimate the SFR of the nearby spiral galaxies, since the
dust-to-gas ratios of these galaxies are almost same as that in the
Galaxy.

\acknowledgments

The author appreciates anonymous referee's great efforts towards
understanding this paper and his many useful suggestions to improve
quality of this work significantly.
The author is most grateful to T. T. Takeuchi for careful reading of
this paper and a lot of useful comments.
The author also thanks very much H. Kamaya and H. Hirashita for their
suggestions and discussions to improve this work.
The author has made extensive use of NASA's Astrophysics Data System
Abstract Service (ADS).

\clearpage

\begin{deluxetable}{lcccc}
\footnotesize
\tablecaption{The Galactic H {\sc ii} regions\label{t1}}
\tablewidth{0pt}
\tablehead{\colhead{Object} & \colhead{$\log N'_{\rm LC}$} & 
\colhead{$\log L_{\rm FIR}$} & \colhead{$L_{\rm IR}/N'_{\rm LC}$} & 
\colhead{$f$} \\
\colhead{} & \colhead{(s$^{-1}$)} & \colhead{($L_\sun$)} & 
\colhead{($10^7L_\sun/10^{50}{\rm s}^{-1}$)} & \colhead{} \\
\colhead{(1)} & \colhead{(2)} & \colhead{(3)} & \colhead{(4)} & 
\colhead{(5)} \\}

 \startdata
 NGC2024 & 47.7 & 4.3 & 0.40 & 0.40 \\
 W3      & 48.2 & 5.3 & 1.26 & 0.14 \\
         & 49.4 & 6.1 & 0.50 & 0.32 \\
 M8      & 48.3 & 4.7 & 0.25 & 0.59 \\
 NGC6357 & 48.4 & 5.2 & 0.63 & 0.26 \\
         & 48.4 & 5.0 & 0.35 & 0.44 \\
 IC4628  & 48.6 & 5.4 & 0.63 & 0.26 \\
 Orion   & 48.7 & 5.2 & 0.35 & 0.44 \\
 G343.4-0.4 & 48.7 & 5.5 & 0.56 & 0.29 \\
 DR15    & 48.8 & 5.8 & 1.12 & 0.15 \\
 G5.9-0.4  & 49.1 & 5.6 & 0.32 & 0.49 \\
 W75/DR21 & 49.2 & 5.3 & 0.13 & 1.01 \\
 RCW117  & 49.5 & 6.1 & 0.35 & 0.44 \\
 G351.6+0.2 & 49.7 & 6.1 & 0.25 & 0.59 \\
 W58     & 49.7 & 6.0 & 0.20 & 0.71 \\
 G351.6-1.3 & 49.8 & 6.3 & 0.32 & 0.49 \\
 W51     & 49.8 & 6.2 & 0.25 & 0.59 \\
         & 50.4 & 6.9 & 0.32 & 0.49 \\
 RCW122  & 49.9 & 6.3 & 0.25 & 0.59 \\
 Sgr C   & 50.0 & 6.8 & 0.63 & 0.26 \\
 M17     & 50.1 & 6.5 & 0.25 & 0.59 \\
 Sgr B   & 50.5 & 7.1 & 0.35 & 0.44 \\
 W49     & 50.7 & 7.3 & 0.35 & 0.44 \\
 \enddata

\tablecomments{Col. (1): Object name. Two separate components in W3,
 W51, and NGC6357 are both listed. Col.(2): LC photon production rates
 estimated from radio observations. Col.(3): Observed FIR
 luminosities. Col.(4): Ratios of IR luminosity to LC photon production
 rates. Col.(5): Estimated $f$ from equation (\ref{eq2-1}) by assuming
 $\epsilon = 1$.}

\end{deluxetable}

\clearpage

\begin{deluxetable}{lcccccc}
\footnotesize
\tablecaption{H {\sc ii} regions in M31\label{t2}}
\tablewidth{0pt}
\tablehead{\colhead{Object ID} & \colhead{$F_{60\micron}$} & 
\colhead{$L_{\rm IR}$} & \colhead{$F_{\rm H\alpha}$} & 
\colhead{$N'_{\rm LC}$} & \colhead{$L_{\rm IR}/N'_{\rm LC}$} & 
\colhead{$f$}\\
\colhead{} & \colhead{(Jy)} & \colhead{($10^6 L_\sun$)} & 
\colhead{($10^{-12}$ erg s$^{-1}$ cm$^{-2}$)} & 
\colhead{($10^{50} {\rm s}^{-1}$)} & 
\colhead{($10^7L_\sun/10^{50}{\rm s}^{-1}$)} & \colhead{} \\
\colhead{(1)} & \colhead{(2)} & \colhead{(3)} & \colhead{(4)} & 
\colhead{(5)} & \colhead{(6)} & \colhead{(7)} \\}

 \startdata
 16 & 3.50 & 7.07 & 3.33 & 3.53 & 0.20 & 0.59 \\
 17 & 3.40 & 6.87 & 1.81 & 1.92 & 0.36 & 0.36 \\
 22 & 4.56 & 9.21 & 4.37 & 4.63 & 0.20 & 0.59 \\
 23 & 4.53 & 9.15 & 2.23 & 2.36 & 0.39 & 0.34 \\
 25 & 3.96 & 8.00 & 0.77 & 0.82 & 0.98 & 0.14 \\
 27 & 2.58 & 5.21 & 1.83 & 1.94 & 0.27 & 0.46 \\
 28 & 6.55 & 13.2 & 3.51 & 3.72 & 0.36 & 0.36 \\
 30 & 1.84 & 3.72 & 0.69 & 0.73 & 0.51 & 0.26 \\
 31 & 6.52 & 13.2 & 2.08 & 2.20 & 0.60 & 0.23 \\
 33 & 7.12 & 14.4 & 3.58 & 3.79 & 0.38 & 0.34 \\
 35 & 2.27 & 4.59 & 1.12 & 1.19 & 0.39 & 0.34 \\
 39 & 2.81 & 5.68 & 2.33 & 2.47 & 0.23 & 0.53 \\
 \enddata

\tablecomments{Col. (1): Identification of 60 \micron\ sources in
 \cite{xu96}. Col.(2): Extracted 60 \micron\ fluxes from {\it IRAS}
 HiRes image by \cite{xu96}. Col.(3): Estimated IR luminosities by
 adopting the distance of 760 kpc. Col.(4): H$\alpha$ fluxes
 corresponding with 60 \micron\ sources. Col.(5): LC photon production
 rates estimated from H$\alpha$ fluxes in col.(4) by adopting the
 distance of 760 kpc and $A_{\rm H\alpha}=0.8$ mag. Col.(6): Ratios of
 IR luminosity to LC photon production rates. Col.(7): Estimated $f$
 from equation (\ref{eq2-1}) when we assume $\epsilon = 0.7$.}

\end{deluxetable}

\clearpage

\begin{deluxetable}{lccccc}
\footnotesize
\tablecaption{H {\sc ii} regions in M33\label{t3}}
\tablewidth{0pt}
\tablehead{\colhead{Object} & \colhead{$L_{\rm FIR}$} & 
\colhead{$F_{\rm 5GHz}$ } & \colhead{$N'_{\rm LC}$} & 
\colhead{$L_{\rm IR}/N'_{\rm LC}$} & \colhead{$f$}\\
\colhead{} & \colhead{($10^7 L_\sun$)} & \colhead{(mJy)} & 
\colhead{($10^{50} {\rm s}^{-1}$)} & 
\colhead{($10^7L_\sun/10^{50}{\rm s}^{-1}$)} & \colhead{} \\
\colhead{(1)} & \colhead{(2)} & \colhead{(3)} & \colhead{(4)} & 
\colhead{(5)} & \colhead{(6)}  \\}

 \startdata
 VGHC 20,25 & 0.56 & 0.8  & 0.50 & 1.12 & 0.13 \\
 IC 131     & 1.00 & 0.7  & 0.44 & 2.27 & 0.064 \\
 IC 133     & 1.67 & 6.6  & 4.14 & 0.40 & 0.33 \\
 NGC 595    & 0.79 & 18.0 & 11.3 & 0.070 & 1.23 \\
 VGHC 46,52 & 1.34 & 1.8  & 1.13 & 1.19 & 0.12 \\
 VGHC 104   & 0.58 & 1.1  & 0.69 & 0.84 & 0.17 \\
 VGHC 97,98 & 0.47 & 4.5  & 2.82 & 0.17 & 0.67 \\
 NGC 604    & 7.50 & 60.0 & 37.6 & 0.20 & 0.59 \\
 \enddata

\tablecomments{Col.(1): Object name. Col.(2): FIR luminosities
 determined from {\it IRAS} HiRes images of 60 and 100 \micron\ by
 adopting the distance of 840 kpc. Col.(3): Observed 5 GHz flux
 densities. Col.(4): LC photon production rates estimated from
 col.(3) by adopting the distance of 840 kpc. Col.(5): Ratios of IR
 luminosity to LC photon production rates. Col.(6): Estimated $f$ from
 equation (\ref{eq2-1}) by assuming $\epsilon = 0.7$.}

\end{deluxetable}

\clearpage

\begin{deluxetable}{lccccccccc}
\footnotesize
\tablecaption{H {\sc ii} regions in LMC\label{t4}}
\tablewidth{0pt}
\tablehead{\colhead{Object} & \colhead{$F_{60\micron}$} & 
\colhead{$F_{100\micron}$} & \colhead{$L_{\rm IR}$} & 
\colhead{$F_{\rm 5GHz}$} & \colhead{$N'_{\rm LC}$} & 
\colhead{$L_{\rm IR}/N'_{\rm LC}$} & \colhead{$E_{B-V}$} & 
\colhead{$\epsilon$} & \colhead{$f$}\\
\colhead{} & \colhead{(Jy)} &\colhead{(Jy)} & 
\colhead{($10^6 L_\sun$)} & \colhead{(Jy)} & 
\colhead{($10^{50} {\rm s}^{-1}$)} & 
\colhead{($10^7L_\sun/10^{50}{\rm s}^{-1}$)} & \colhead{(mag)} & 
\colhead{} & \colhead{} \\
\colhead{(1)} & \colhead{(2)} & \colhead{(3)} & \colhead{(4)} & 
\colhead{(5)} & \colhead{(6)} & \colhead{(7)} & \colhead{(8)} & 
\colhead{(9)} & \colhead{(10)}\\}

 \startdata
 MC 18    & 1229 & 2029 & 8.48 & 3.37 & 8.03 & 0.11 & 0.07 & 0.28 &
 0.68 \\
 MC 47    & 307  & 363  & 1.88 & 0.64 & 1.53 & 0.12 & 0.14 & 0.47 &
 0.72 \\
 MC 57    & 337  & 509  & 2.25 & 0.81 & 1.93 & 0.12 & 0.13 & 0.45 &
 0.74 \\
 MC 64    & 524  & 826  & 3.55 & 1.65 & 3.93 & 0.090 & 0.24 & 0.66 &
 1.03 \\
 MC 71    & 594  & 819  & 3.83 & 0.55 & 1.31 & 0.29 & 0.11 & 0.39 &
 0.34 \\
 MC 90+91 & 204  & 382  & 1.48 & 0.87 & 2.07 & 0.071 & 0.10 & 0.37 &
 0.95 \\
 \enddata

\tablecomments{Col.(1): Object name. Cols.(2) and (3): {\it IRAS} co-added
 flux densities at 60 and 100 $\mu$m, respectively. Col.(4): IR luminosities
 determined from cols.(2) and (3). The adopted distance is 51.8
 kpc. Col.(5): Observed 5 GHz flux densities. Col.(6): LC photon
 production rates estimated from col.(5) when the distance is adopted as 
 51.8 kpc. Col.(7): Ratios of IR luminosity to LC photon production
 rates. Col.(8): $E_{B-V}$ estimated from the observed Balmer
 decrement. Col.(9): $\epsilon$ determined from Figure
 \ref{f1}. Col.(10): Estimated $f$ from equation (\ref{eq2-1}).}

\end{deluxetable}

\clearpage

\begin{deluxetable}{lccc}
\footnotesize
\tablecaption{Summary of determined $f$ for sample galaxies\label{t5}}
\tablewidth{0pt}
\tablehead{\colhead{Galaxy} & \colhead{$f$} & 
\colhead{12+log(O/H)} & \colhead{$\cal D$/$\cal D_{\rm MW}$} \\
\colhead{(1)} & \colhead{(2)} & \colhead{(3)} & \colhead{(4)} \\}

 \startdata
 Galaxy    & $0.45 \pm 0.21$ & 8.7 & 1.0 \\
 M31       & $0.38 \pm 0.14$ & 9.0 & 1.2 \\
 M33       & $0.41 \pm 0.37$ & 8.4 & 0.6 \\
 LMC       & $0.74 \pm 0.22$ & 8.37& 0.2 \\
\enddata

\tablecomments{Col.(1): Galaxy name. Col.(2): Mean $f$ with the sample
 standard deviation. Col.(3): Metallicity from \cite{van00}. Col.(4):
 Dust-to-gas mass ratio normalized by the Galactic value, 
 $6 \times 10^{-3}$.}

\end{deluxetable}

\clearpage

\appendix

\section{Infrared excess (IRE)}

The IR excess (IRE) is often examined for H {\sc ii} regions.
We examine the relation between the LCE and IRE.
Especially, we relate $f$ to IRE directly here.
A concerned discussion is found in \cite{hir01}.

\begin{figure}
\plotone{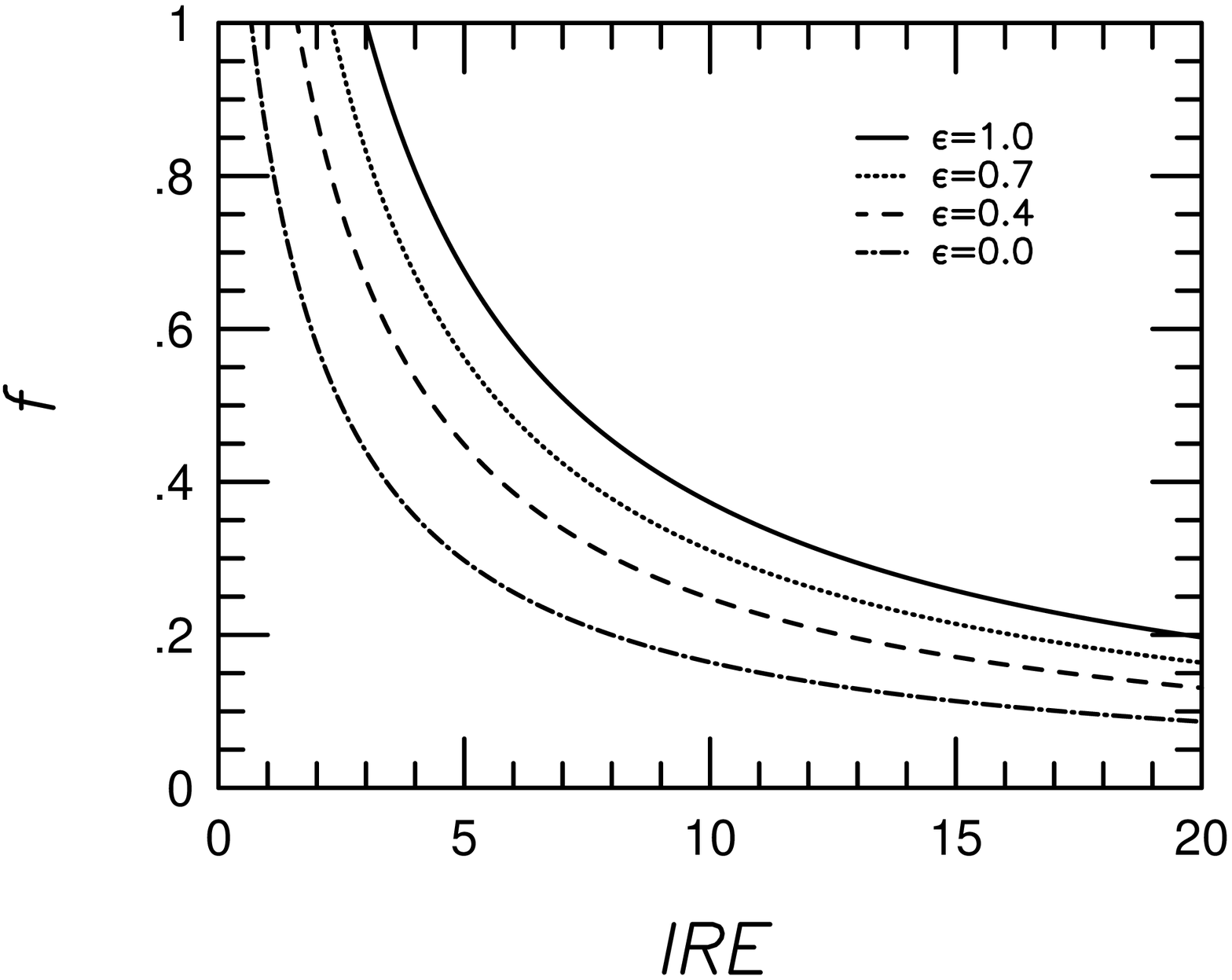}
\figcaption{Fraction of Lyman continuum photons contributing to
hydrogen ionization, $f$ as a function of infrared excess, $IRE$. The
solid, dotted, dashed, and dash-dotted lines are calculated by equation
(\ref{eqa-2}) with $\epsilon =$ 1.0, 0.7, 0.4, and 0,
respectively.\label{fa}}
\end{figure}

Mezger et al.\ (1974) have defined the IRE as 
\begin{equation}
 IRE \equiv \frac{L_{\rm IR}}{N'_{\rm LC}h \nu_{\rm Ly \alpha}}\,,
 \label{eqa-1}
\end{equation}
where $h$ is the Plank constant and $\nu_{\rm Ly \alpha}$ denotes the
frequency at the Lyman $\alpha$ line.
{}From equation (\ref{eqa-1}), we obtain $L_{\rm IR,7}/N'_{\rm LC,50} =
4.23 \times 10^{-2} IRE$.
Therefore, equation (\ref{eq2-1}) is reduced to
\begin{equation}
 f = \frac{0.44 + 0.56 \epsilon}{0.28 + 0.24 IRE}\,.
 \label{eqa-2}
\end{equation}
In Figure \ref{fa}, we show the relation between $f$ and $IRE$ derived
from equation (\ref{eqa-2}) for various $\epsilon$.

The solid, dotted, and dashed lines correspond to $\epsilon = 1.0$ (in
the Galaxy), 0.7 (in M31 and M33), and 0.4 (in LMC), respectively.
In addition, we also present the dash-dotted line of $\epsilon = 0$ for 
comparison, which corresponds to the case that no nonionizing photons
are absorbed by dust.
According to \cite{spi78}, the probability of producing Lyman $\alpha$
photons is two-thirds per every ionization--recombination process of
hydrogen.
Thus, when the observed IR luminosity of an H {\sc ii} region is
explained by only the luminosity of Lyman $\alpha$ photons produced in
the region ($f=1$ and $\epsilon=0$), $IRE$ will be 0.67.

\cite{ant87} have reported $4 < IRE < 40$ for the Galactic
compact H {\sc ii} regions.
The mean IRE of their sample is about 12.
For such high IRE regions, we expect that $f$ is about 0.3 or less.
When we estimate the SFR of such regions, therefore, we should not use
the H$\alpha$ or thermal radio luminosities but use the IR luminosity as
an indicator of the SFR.

\end{document}